\documentclass{vgtc}                          

\ifpdf
  \pdfoutput=1\relax                   
  \usepackage{graphicx}                
  \DeclareGraphicsExtensions{.pdf,.png,.jpg,.jpeg} 
\else
  \usepackage{graphicx}                
  \DeclareGraphicsExtensions{.eps}     
\fi%

\graphicspath{{figures/}{pictures/}{images/}{./}} 

\usepackage{microtype}                 
\PassOptionsToPackage{warn}{textcomp}  
\usepackage{textcomp}                  
\usepackage{mathptmx}                  
\usepackage{times}                     
\usepackage{cite}                      
\usepackage{tabu}                      
\usepackage{booktabs}                  

\onlineid{1405}

\vgtccategory{Methodological Paper}

\vgtcinsertpkg

\title{Construction of a Validated Virtual Embodiment Questionnaire}

\author{Daniel Roth\thanks{e-mail: daniel.roth@uni-wuerzburg.de}\\ %
        \scriptsize HCI Group, University of Würzburg %
\and Marc Erich Latoschik\thanks{e-mail: marc.latoschik@uni-wuerzburg.de}\\ %
     \scriptsize HCI Group, University of Würzburg %
}

\teaser{
	\centering
	\includegraphics[width=.85\linewidth]{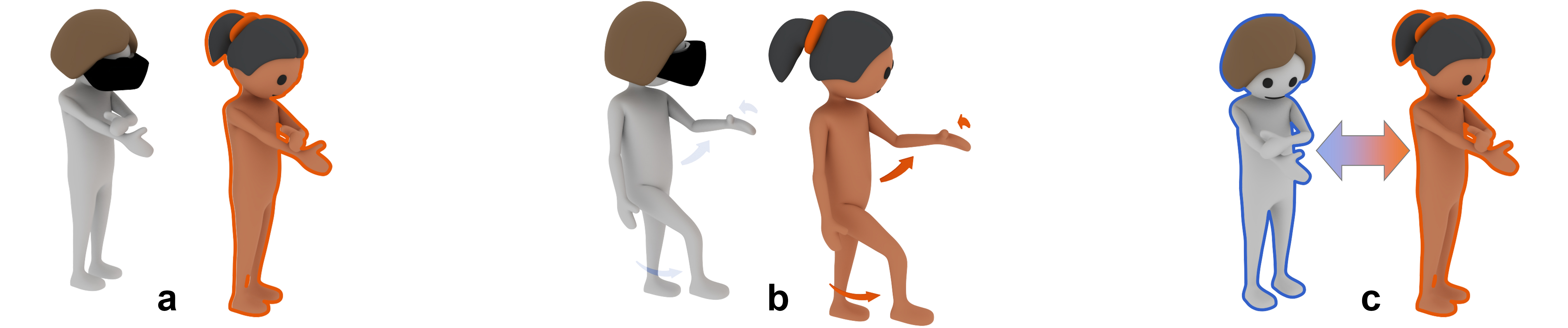}
	\vspace{-2ex}
	\caption[Virtual body ownership, agency, change.]{\textbf{Factors addressed by the VEQ.} Virtual body ownership (a), agency (b), and the change in the perceived body schema (c). The user is depicted in grey; the virtual avatar is depicted in orange.}
	\vspace{-1ex}
	\label{fig:comp-scale}
}

\abstract{User embodiment is important for many virtual reality (VR) applications, for example, in the context of social interaction, therapy, training, or entertainment. However, there is no validated instrument to empirically measure the perception of embodiment, necessary to reliably evaluate this important quality of user experience (UX). To assess components of virtual embodiment in a valid, reliable, and consistent fashion, we develped a Virtual Embodiment Questionnaire (VEQ). We reviewed previous literature to identify applicable constructs and items, and performed a confirmatory factor analysis (CFA) on the data from three experiments ($N=196$). Each experiment modified a distinct simulation property, namely, the level of immersion, the level of personalization, and the level of behavioral realism. The analysis confirmed three factors: (1) ownership of a virtual body, (2) agency over a virtual body, and (3) change in the perceived body schema. A fourth study ($N=22$) further confirmed the reliability and validity of the scale and investigated the impacts of latency jitter of avatar movements presented in the simulation compared to linear latencies and a baseline. We present the final scale and further insights from the studies regarding related constructs.}

\CCScatlist{
  \CCScatTwelve{Human-centered computing}{Visu\-al\-iza\-tion}{Visualization design and evaluation methods}{}
}

 \nocopyrightspace

\begin{document}


\firstsection{Introduction}

\maketitle


User embodiment can be referred to as ``the provision of users with appropriate
body images so as to represent them to others (and also to themselves)'' \cite[p.242]{benford_user_1995}. With the availability of consumer VR technology, it became apparent that the relevance of virtual embodiment is not limited to the understanding of cognitive processes. In addition, understanding virtual embodiment has concrete and direct implications for research, design, and development. It is especially relevant for applications that consider therapy \cite{rizzo_virtual_2009,muhlberger_repeated_2001,anderson_virtual_2013,riva_treating_1997,dallaire-cote_animated_2016,piryankova_owning_2014}, entertainment \cite{ratan_leveling_2015,klevjer_enter_2012,lugrin2018any}, as well as collaboration and social interaction \cite{schroeder_avatars_2006,kulik_virtual_2018,beck_immersive_2013,latoschik_effect_2017}. Previous research investigated virtual embodiment in various studies. However, a consistent and validated instrument for assessing components of virtual embodiment, to the best of our knowledge, does not exist.

\vspace{3ex}
\textit{© 20xx IEEE. Personal use of this material is permitted. Permission from IEEE must be obtained for all other uses, in any current or future media, including reprinting/republishing this material for advertising or promotional purposes, creating new collective works, for resale or redistribution to servers or lists, or reuse of any copyrighted component of this work in other works.}

\subsection{The Demand for a Standardized Measure}
Previous studies adapted measures from originating experiments such as the rubber hand illusion (RHI) \cite{botvinick_rubber_1998}, for example, individual questionnaires and displacement measures. However, the assessment of concepts such as virtual body ownership (VBO) is not consistent. Effects are often assessed with single items, which was argued to be problematic \cite{carifio_ten_2007}. Approaches to cross-validating different measures often failed (see \cite{kilteni_sense_2012,roth_alpha_2017} for discussions). 
According to Boateng et al. \cite{boateng_best_2018}, the creation of a rigorous scale undergoes three stages: 1) In the \textit{item development stage}, the domain is identified, and items are theoretically analyzed regarding their content validity. 2) In the \textit{scale development} stage, questions are pre-tested, and hypothesized factors are explored with covariance analysis (e.g., exploratory factor analysis), along with a consecutive reduction of items to an item pool that remains internally consistent, followed by the factor extraction. In the \textit{scale evaluation} phase, the dimensionality is tested with CFA, and consecutively, the reliability (e.g., Cronbach's alpha) and validity (i.e., relation to other constructs and measurements) are evaluated.  

Gonzalez-Franco and Peck emphasized the request for a standardized questionnaire and review assessments \cite{gonzalez-franco_avatar_2018}. From those, they ``identify a set of questions \textit{to be standardized} for future embodiment experiences'' \cite[p. 4]{gonzalez-franco_avatar_2018} which they organized in six experimental interests (body ownership, agency and motor control, tactile sensations, location of the body, external appearance, and response to external stimuli). 
Progressing beyond a theoretical approach, we propose a scale that is data driven, and proceeded through all stages of rigorous scale development (see also \cite{boateng_best_2018,furr2011scale}), and is generally applicable to many embodiment experiments.
A scale should be \textbf{reliable} (produce repeatable results within and across subjects irrespective of the testing conditions), \textbf{valid} (measure the underlying constructs precisely), \textbf{sensitive} (discriminate between multiple outcome levels), and \textbf{objective} (shielded from third-party variable bias) \cite{witmer_measuring_1998,meehan2002physiological}.

\subsection{Contribution}
We present the creation of a valid and consistent measurement instrument for assessing three components of virtual embodiment (ownership, agency, and change in perceived body scheme). These components were explored, confirmed, and validated. We investigated the scale performance by exploring it's relation to related concepts and previous measures applied, and confirm its validity, reliability, and sensitivity. We further provide insights into the impacts of immersion, avatar personalization, behavioral realism, and latency and latency jitter that were investigated with the experiments. The questionnaire can be applied to various VR experiments to assess virtual embodiment.

\subsection{Embodiment}


Embodiment is a part of self-consciousness and arises through multisensory information processing \cite{lopez_body_2008,lenggenhager_video_2007,gallagher_how_2006}. Previous literature mainly considered three components of embodiment: A conscious experience of self-identification (body ownership), controlling one's own body movements (agency) \cite{tsakiris_having_2006}, and being located at the position of one's body in an environment (self-location) \cite{lopez_body_2008,lenggenhager2009spatial}. 
Research also stresses the importance of the perspective with which one perceives the world with (first-person perspective) \cite{blanke_full-body_2009,blanke_multisensory_2012}. 

\textbf{Body ownership} can be described as the experience and allocation of a bodily self as one's own body, as ``my body,'' the particular perception of one's own body as the source of bodily sensations, unique to oneself so that it is present in one's mental life \cite{gallagher_philosophical_2000,tsakiris_having_2006,tsakiris_my_2010}. A key instrument in investigations of body ownership is the RHI \cite{botvinick_rubber_1998}. The experiment stimulates ownership of an artificial body part in the form of a rubber hand by simultaneous tactile stimulation (visually hidden) of the physical hand combined with a parallel visual stimulation of a rubber hand. Caused by the stimulation, participants start to perceive the rubber hand as part of their body. 

\textbf{Agency}, meaning the ``experience of oneself as the agent of one's own actions - and not of others' actions'' (\cite[p. 523,]{david_sense_2008} following \cite{gallagher_philosophical_2000}) relates to body ownership \cite{tsakiris_having_2006,braun_senses_2018,tsakiris_agency_2007}. Tsakiris, et al. described agency as ``the sense of intending and executing actions, including the feeling of controlling one's own body movements, and, through them, events in the external environment'' \cite[p. 424]{tsakiris_having_2006}. 

Previous work showed that bottom-up accounts (multisensory processing and integration) are an important driver \cite{kilteni_over_2015,tsakiris_my_2010}, and that top-down processes (e.g., form and appearance matching) at least modulate embodiment \cite{haans_effect_2008,tsakiris_having_2006,braun_senses_2018,tsakiris_agency_2007}. 
Kilteni, et al. \cite{kilteni_over_2015} reviewed triggers and preventers of body ownership illusions and summarized that cross-modal stimulation, for example, congruent visuomotor and visuotactile stimulation supported ownership illusions. In contrast, incongruencies (thus counter-acting sensorimotor contingencies) hinder ownership. Further, visuoprorioceptive cues, such as perspective shifts or modified distances, as well as semantic modulations can impact ownership illusions \cite{kilteni_over_2015}.


\subsection{Virtual Embodiment}
\label{sec:vbo}

%
%

Virtual embodiment was defined as ``the physical process that employs the VR hardware and software to substitute a person's body with a virtual one'' \cite[p.1]{spanlang2014build}. Embodiment has received ongoing attention in VR research, for example, regarding avatar hand appearance \cite{argelaguet_role_2016,schwind_wheres_2017,jung2018over,hoyet_wow!_2016} and full body representations \cite{slater_first_2010,maselli_building_2013,waltemate_impact_2018,mohler2010effect,piumsomboon_mini-me_2018}. Avatars that represent a user are defined as virtual characters driven by human behavior \cite{bailenson_avatars_2004}. Following video-based approaches \cite{lenggenhager_video_2007}, researchers found that the concept of the RHI also applies to virtual body parts \cite{slater_towards_2008}, and entire virtual bodies \cite{slater_inducing_2009,slater_first_2010,lugrin_anthropomorphism_2015}. 

To investigate and alternate virtual body perception, experimentation has used mirrors in immersive HMD-based simulations \cite{gonzalez-franco_contribution_2010}, and semi-immersive fake (magic) mirror projections \cite{latoschik_fakemi_2016,waltemate_impact_2018}. According to Slater, et al. \cite{slater_inducing_2009}, the induction time for body ownership illusions varies between about 10 s and 30 min. Kilteni, et al. \cite{kilteni_sense_2012} summarized findings and measures regarding self-location, agency, and body ownership, and argue for a continuous measurement approach. Similar to Kilteni, et al. \cite{kilteni_over_2015}, Maselli and Slater concluded from their experiments that bottom-up factors, like sensorimotor coherence, and a first-person perspective, are driving factors \cite{maselli_building_2013}. They argue that appearance moderates the experience insofar as realistic humanoid textures foster body ownership. 

An important effect regarding the embodiment through avatars in VR is the Proteus effect \cite{yee_walk_2006,yee_proteus_2007}. Yee and Bailenson found a change in behavior, self-perception and participants' identity when taking the perspective of an avatar with altered appearance. Participants changed their behavior and attitude according to behavior they attributed to their virtual representation. 
 
In summary, the degree and the precision with which sensory stimulation, appearance, and behavior are rendered, as well as the perspective that is presented are important aspects for user-embodying interfaces \cite{maselli_building_2013,kilteni_over_2015,lugrin_anthropomorphism_2015,gonzalez-franco_contribution_2010,waltemate_realizing_2015,waltemate_impact_2018}. 
Design choices, for example, the character type, or the realism of the replicated appearance and behavior may strongly influence the perceptual phenomena of embodiment. Measuring embodiment in a valid way, therefore, is crucial to VR applications.

\section{Scale Construction}
\label{sec:intra:beta-gold}

\begin{table}[tb]
	\caption[Proposed items for generalization and extension.]{Proposed items for the generalization and extension.}
	\vspace{-6ex}
	\label{tab:betaivbo}
	\scriptsize
	\begin{center}
		\begin{tabular}{l}
			\midrule
			\textbf{\textit{Ownership}} \\ 
			- It felt like the virtual body was my body. \\ \vspace{-2.5ex}
			
			\\
			- It felt like the virtual body parts were my body parts. \\ \vspace{-2.5ex}

			\\
			- The virtual body felt like a human body. \\ \vspace{-2.5ex}
			
			\\  - I had the feeling that the virtual body belonged to another person.\textbf{*}$\dagger$
			\\  \vspace{-2.5ex}
			\\ 
			\vspace{-2.5ex} - It felt like the virtual body belonged to me.$\dagger$
			\\				
			\\
			\textbf{\textit{Agency}} \\
			- The movements of the virtual body seemed to be my own movements. \\    \vspace{-2.5ex}
			
			\\
			- I enjoyed controlling the virtual body.	\\ \vspace{-2.5ex}
			
			\\
			- I felt as if I was controlling the movement of the virtual body. \\  \vspace{-2.5ex}

			\\ 
			- I felt as if I was causing the movement  of the virtual body.	\\   \vspace{-2.5ex}

			\\ 
			- The movements of the virtual body were synchronous with my \\ \vspace{-2.5ex} own movements.$\dagger$
			\\ 
			
			\\ 
			\textbf{\textit{Change (in the perceived body schema)}} \\
			- I had the illusion of owning a different body from my own (body).  \\   \vspace{-2.5ex}		
			
			\\  
			- I felt as if the form or appearance of my body had changed.
			\\    \vspace{-2.5ex}
			
			\\  
			- I felt the need to check if my body really still looked like what \\ I had in mind.
			\\ \vspace{-2.5ex}
			
			\\	
			- I had the feeling that the weight of my body had changed.
			\\  \vspace{-2.5ex}
			
			\\  
			- I had the feeling that the height of my body had changed.
			\\   \vspace{-2.5ex}
			
			\\  
			- I had the feeling that the width of my body had changed.\\
			\midrule 
			\vspace{-3ex}\\
			
			\em{Note. * Required recoding, $\dagger$ new item.}
			\vspace{-7ex}
			
		\end{tabular} 
	\end{center}
\end{table}

Previous assessments often base on the original RHI experiment \cite{botvinick_rubber_1998}. A psychometric approach to assessing levels of embodiment toward an artificial physical body part identified the latent variables of ownership, agency, and location \cite{longo_what_2008}. The assessment of location included items focusing on the coherence between sensation and causation, and locational similarities of artificial physical body parts. 

Roth et al. \cite{roth_alpha_2017} presented a proposal to assess virtual embodiment with 13 questions from previous work \cite{lugrin_anthropomorphism_2015,botvinick_rubber_1998,slater_first_2010,steptoe_human_2013,petkova_if_2008,ehrsson_experimental_2007,armel2003projecting,kalckert_moving_2012,gonzalez-franco_contribution_2010,powers_advisor_2006,longo_what_2008}. A principal component analysis revealed three factors: acceptance (covering the aspect of ownership perception), control (covering the aspect of agency perception), and change (covering the aspect of a perceived change in one's own body schema). The last factor may be especially important for studies that make use of altered body appearances and the Proteus effect \cite{yee_proteus_2007,yee_walk_2006,roth_alpha_2017}. The instrument showed good reliability in further assessments \cite{waltemate_impact_2018,latoschik_effect_2017}. In addition to necessary validation and consistency analysis, two downsides can be identified: 1) The measure is strongly constrained to ``virtual mirror'' scenarios due to the phrasing, and 2) the components are not balanced regarding the number of items. In particular, the ownership component consists of only three statements.
We based our scale construction on their work and performed necessary improvements. Regarding the item development, we first generalized the phrasing of the questions to fit generic scenarios. Second, we added questions to balance the component assessment for \textit{ownership} (``It felt like the virtual body belonged to another person,'' - adapted from \cite{petkova_if_2008}, ``It felt like the virtual body belonged to me'') and \textit{agency} factors (``The movements of the virtual body were in sync with my own movements''), respectively. 
The items are shown in Table \ref{tab:betaivbo}. For the \textit{scale development} and \textit{scale evaluation}, CFA was calculated with the data from three studies ($N=196$) that assessed the impacts of particular manipulations hypothesized to affect embodiment. The study results are discussed in detail in Section \ref{sec:studies}.

%
%
%
%
%
%
%
%

\subsection{Confirmatory Factor Analysis}
\label{sec:intra:cfa}	
We performed the CFAs using R with the lavaan package. The reporting of fit indices is based on the recommendation made by Kline \cite{kline_principles_2010}. As the assumption of multivariate normality was violated, we conducted a robust maximum likelihood estimation and computed Satorra-Bentler (SB) corrected test statistics (see \cite{brown_confirmatory_2014}). The first attempt did not yield an acceptable model fit, and as the modification indices indicated, there were covariations in the error terms of a particular item loading on several factors (``I had the illusion of owning a different body from my own''). Therefore, this item was dropped. A second attempt yielded an acceptable model fit. However, inspection of the modification indices revealed covariations in the error terms of two items loading on several factors (``I had the feeling that the virtual body belonged to another person,'' ``I enjoyed controlling the virtual body''). Thus, we excluded problematic items. Furthermore, items with a factor loading $<.40$ were excluded (``I felt the need to check if my body really still looked like what I had in mind''). The third CFA with the remaining 16 items yielded a more parsimonious solution with a good model fit, SB $\chi^2= 52.50$, \textit{df}$=51$, $p=.416$,  root mean square error of approximation AC (RMSEA) $= .013$, $90\%$ confidence interval of robust root mean square error of approximation $[.000; .052]$, standardized root mean square residual (SRMR) $= .047$, and a robust comparative fit index (CFI) $= .998$. Thus, the solution was deemed acceptable to characterize these components of virtual embodiment. Table \ref{tab:factorloadings} depicts the standardized coefficients. The reliability values for \textit{ownership} ($\alpha=.783$), agency ($\alpha=.764$), and change ($\alpha=.765$) were acceptable. The resulting scale is depicted in Table \ref{tab:goldivbo}, its factors are illustrated in Fig \ref{fig:comp-scale}. The scale was assessed in german. A professional service was consulted for the translation.

\begin{table}[tb]
	\caption[Factor Loadings.]{The confirmed embodiment factors (CFA results).}
	\vspace{-4ex}
	\label{tab:factorloadings}
	\scriptsize
	\begin{center}
		\begin{tabular}{lccc}

			\midrule \vspace{-.5ex}
				&	Ownership	&	Agency	&	Change \\ 
			\midrule 
			myBody 	&	.81	&		&		\\
			myBodyParts	&	.73	&		&		\\
			human	&	.53	&		&		\\
			belongsToMe	&	.71	&		&		\\
			myMovement	&		&	.80	&		\\
			controlMovements	&		&	.70	&		\\
			causeMovements	&		&	.72	&		\\
			syncMovements	&		&	.53	&		\\
			myBodyChange	&		&		&	.69	\\
			echoHeavyLight	&		&		&	.78	\\
			echoTallSmall	&		&		&	.48	\\
			echoLargeThin	&		&		&	.72	\\

			\midrule 
			\vspace{-3ex}\\
			
			\multicolumn{4}{l}{\em{Note. Coefficients represent standardized path coefficients of the CFA.}} \\  
			\multicolumn{4}{l}{\em{Correlations between factors: Ownership $\sim$ Agency $r = .69$,}} \\
			\multicolumn{4}{l}{\em{Ownership $\sim$ Change $r = .17$, Change $\sim$ Agency $r = -.16$.}}
			\vspace{-7ex}
		\end{tabular} 
	\end{center}
\end{table}

%

\begin{table}[tb]
	\caption[The resulting embodiment questionnaire.]{The Resulting Virtual Embodiment Questionnaire (VEQ).}
	\vspace{-4ex}
	\label{tab:goldivbo}
	\scriptsize
	\begin{center}
		\begin{tabular}{l}
			\midrule 			
			\textbf{\textit{Ownership}} - Scoring: ([OW1] + [OW2] + [OW3] + [OW4]) $\slash$ 4 \\ 
			\textbf{OW1. myBody} \\
			It felt like the virtual body was my body.
			
			\\ \vspace{-2.5ex}
			\\  \textbf{OW2. myBodyParts  }
			\\ \vspace{-2.5ex} It felt like the virtual body parts were my body parts.
			
			\\
			\\  \textbf{OW3. humanness}
			\\ \vspace{-2.5ex} The virtual body felt like a human body.

			\\
			\\  \textbf{OW4. belongsToMe }	
			\\ \vspace{-2.5ex} It felt like the virtual body belonged to me.
			
			\\ \vspace{.5ex}
			\\ \textbf{\textit{Agency}} - Scoring: ([AG1] + [AG2] + [AG3] + [AG4]) $\slash$ 4 \\ 
			\textbf{AG1. myMovement }  
			\\  \vspace{-2.5ex} The movements of the virtual body felt like they were my movements.
			
			\\
			
			\\   \textbf{AG2. controlMovements} 
			\\  \vspace{-2.5ex} I felt like I was controlling the movements of the virtual body.

			\\
			\\   \textbf{AG3. causeMovements } 
			\\ \vspace{-2.5ex} I felt like I was causing the movements of the virtual body.
			
			\\
			\\  \textbf{AG4. syncMovements } 
			\\  	The movements of the virtual body were in sync with my \\ own movements.
			
			\\ \vspace{-2ex}
			\\ \textbf{\textit{Change}} - Scoring: ([CH1] + [CH2] + [CH3] + [CH4]) $\slash$ 4  
			\\ \textbf{CH1. myBodyChange}  
			\\ \vspace{-2.5ex} I felt like the form or appearance of my own body had changed.
			
			\\
			\\   \textbf{CH2. echoHeavyLight}  
			\\  \vspace{-2.5ex} I felt like the weight of my own body had changed.

			\\ 
			\\  \textbf{CH3. echoTallSmall}  
			\\  \vspace{-2.5ex} I felt like the size (height) of my own body had changed.
			
			\\
			\\  \textbf{CH4. echoLargeThin } 
			\\   I felt like the width of my own body had changed.\\
			\midrule 
			\vspace{-3ex}\\
			
			{\em{{Note.} {Participant instructions:} Please read each statement and answer}}\\ 
			{\em{on a 1 to 7 scale indicating how much each statement applied to you}}\\
			{\em{during the experiment. There are no right or wrong answers. Please}}\\
			{\em{answer spontaneously and intuitively. Scale example: 1--strongly}}\\ {\em{disagree, 4--neither agree nor disagree, 7--strongly agree.}}\\
			{\em{A professional service was consulted for the translation.}}\\
			\vspace{-8ex}	
		\end{tabular} 
	\end{center}
\end{table}

\section{Validation}
\label{sec:studies}
The validation of the VEQ is based on four studies that explored individual aspects of virtual embodiment.
We manipulated simulation properties that were hypothesized to affect the perceived embodiment, namely, immersion (Study 1), personalization (Study 2), behavioral realism (Study 3), simulation latency and latency jitter (Study 4). To assess the scale validity, we investigated correlations to related concepts, reliability, and compared the scale performance to related instruments. Participants were individually recruited (i.e., the studies were not performed in block testing fashion) through the recruitment system of the institute for human-computer-media at the University of W\"urzburg. All studies were approved by the ethics committee of the institute for human-computer-media at the University of W\"urzburg.

\subsection{Study 1: Impacts of Immersion}
\label{sec:beta-immersion}
Virtual embodiment was previously assessed in immersive head-mounted display (HMD)-based systems in first- and third-person perspective (e.g., \cite{slater_first_2010}), and with less immersive L-Shape displays \cite{waltemate2016impact}, or fake mirror scenarios \cite{latoschik_fakemi_2016}. Recent work confirmed that visual immersion (i.e., first-person perspective and HMD) fosters virtual embodiment \cite{waltemate_impact_2018}. Thus, we hypothesized H1.1: \textit{A lower immersive display setup is inferior in inducing virtual embodiment}. According to the previous literature \cite{slater_note_2003,slater_place_2009,kilteni_sense_2012}, we also assumed that H1.2: \textit{Higher immersion results in higher presence}. The latter was also assessed to investigate the relation between the VEQ factors and presence. 

\begin{figure*}
	\centering
	\frame{\includegraphics[height=1.53in]{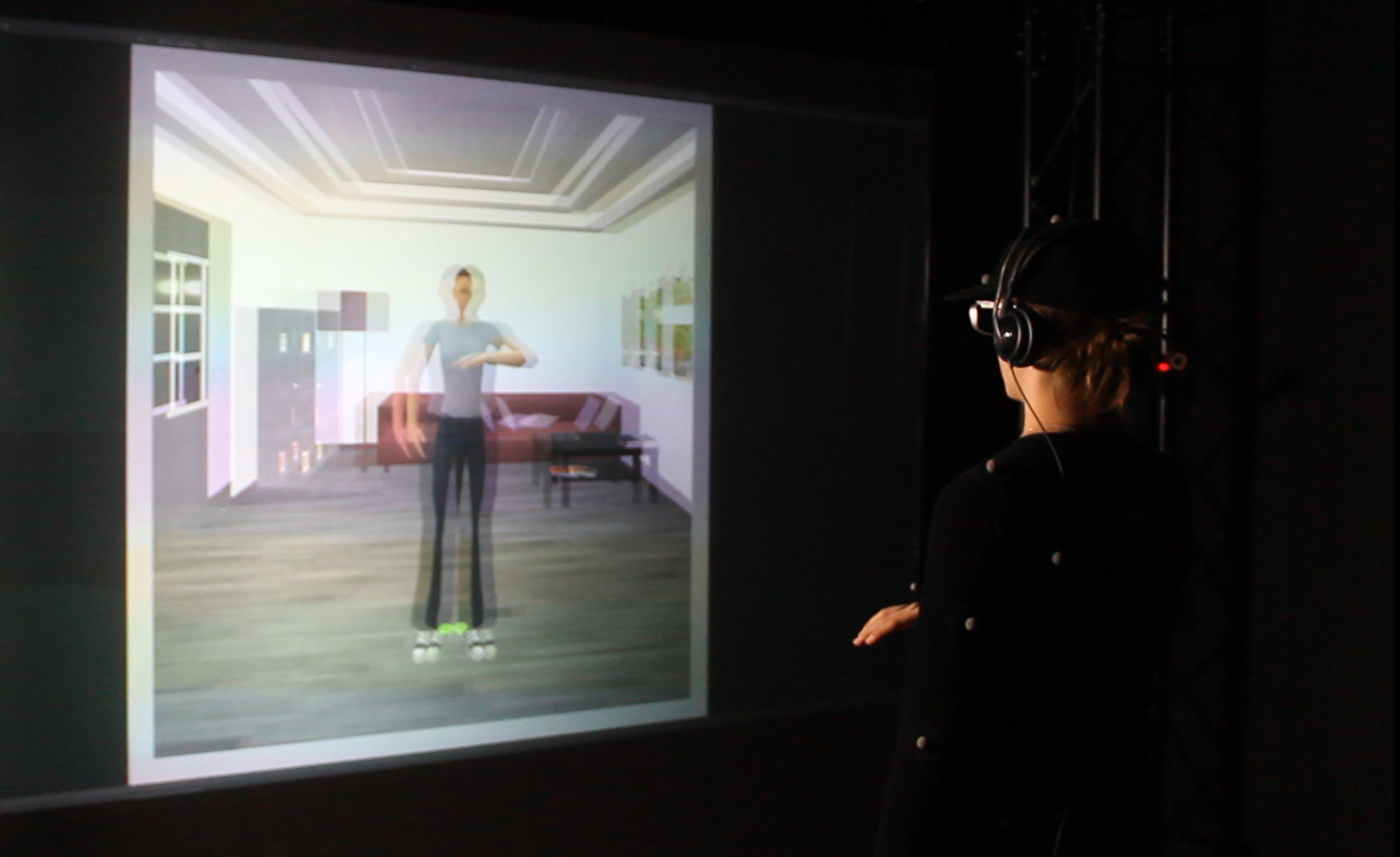}} \hfill
	\frame{\includegraphics[height=1.53in]{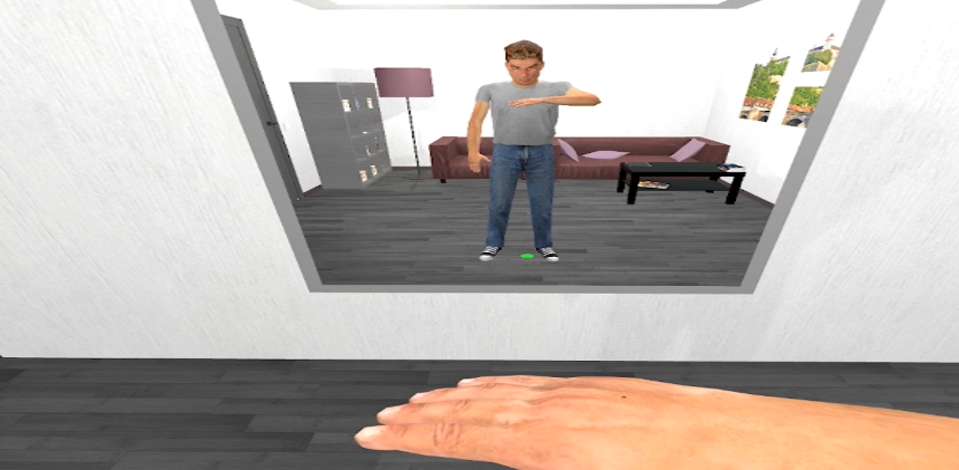} \hspace{-1.1ex}
	\includegraphics[height=1.53in]{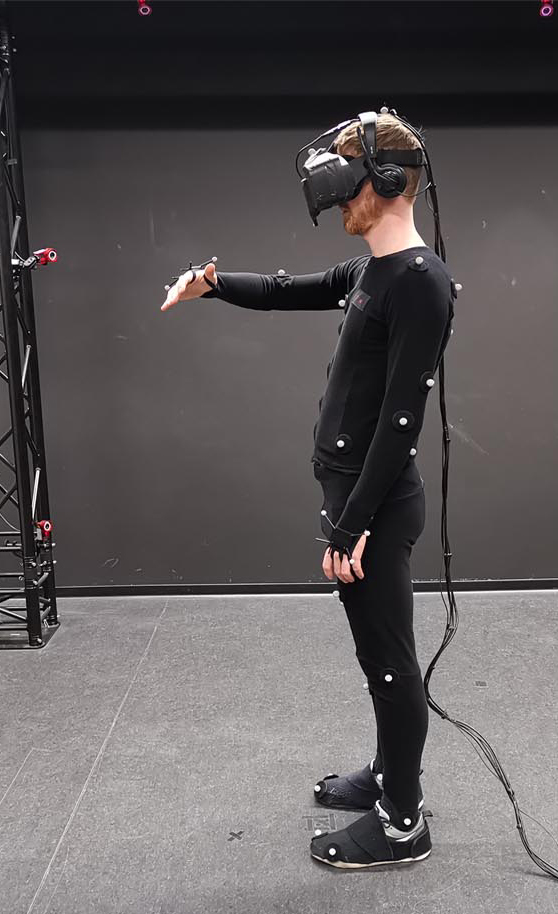}} \hfill
	\frame{\includegraphics[height=1.53in]{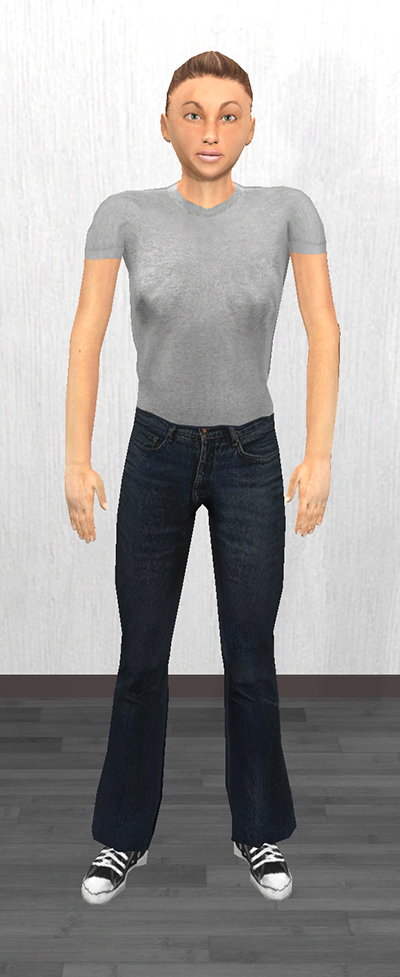} \hspace{-1.1ex}
	\includegraphics[height=1.53in]{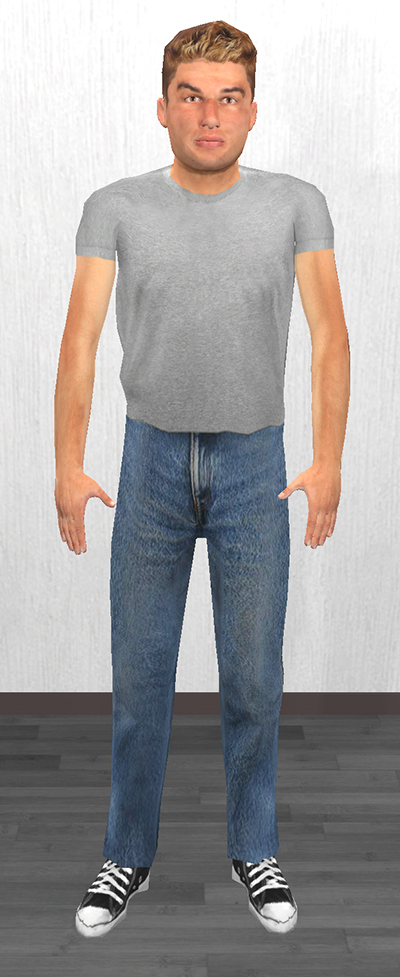}}
	\vspace{-2.5ex}
	\caption[Apparatus of Study 1: Impact of immersion.]{\textbf{Apparatus (Study 1).} \textit{Left:} The projection condition. \textit{Center:} The HMD condition. \textit{Right:} The female and male avatars used for Study 1.}
	\vspace{-2ex}
	\label{fig:immersion-apparatus}
\end{figure*}

\subsubsection{Method}
In a one-factor (\textit{Medium}) between-subjects design modifying the level of immersion, participants were exposed to either a fake-mirror projection, in which case the participant had a visual reference to her or his physical body, or an immersive simulation displayed with an HMD, in which case the participant saw her or his virtual body from a first-person perspective (Fig. \ref{fig:immersion-apparatus}). To provoke the perception of virtual embodiment, participants were asked to perform motions and focus their attention, similar to related experiments \cite{latoschik_fakemi_2016,waltemate_impact_2018}, as described in the following.

\begin{figure}
	\centering
	\includegraphics[width=.6\columnwidth]{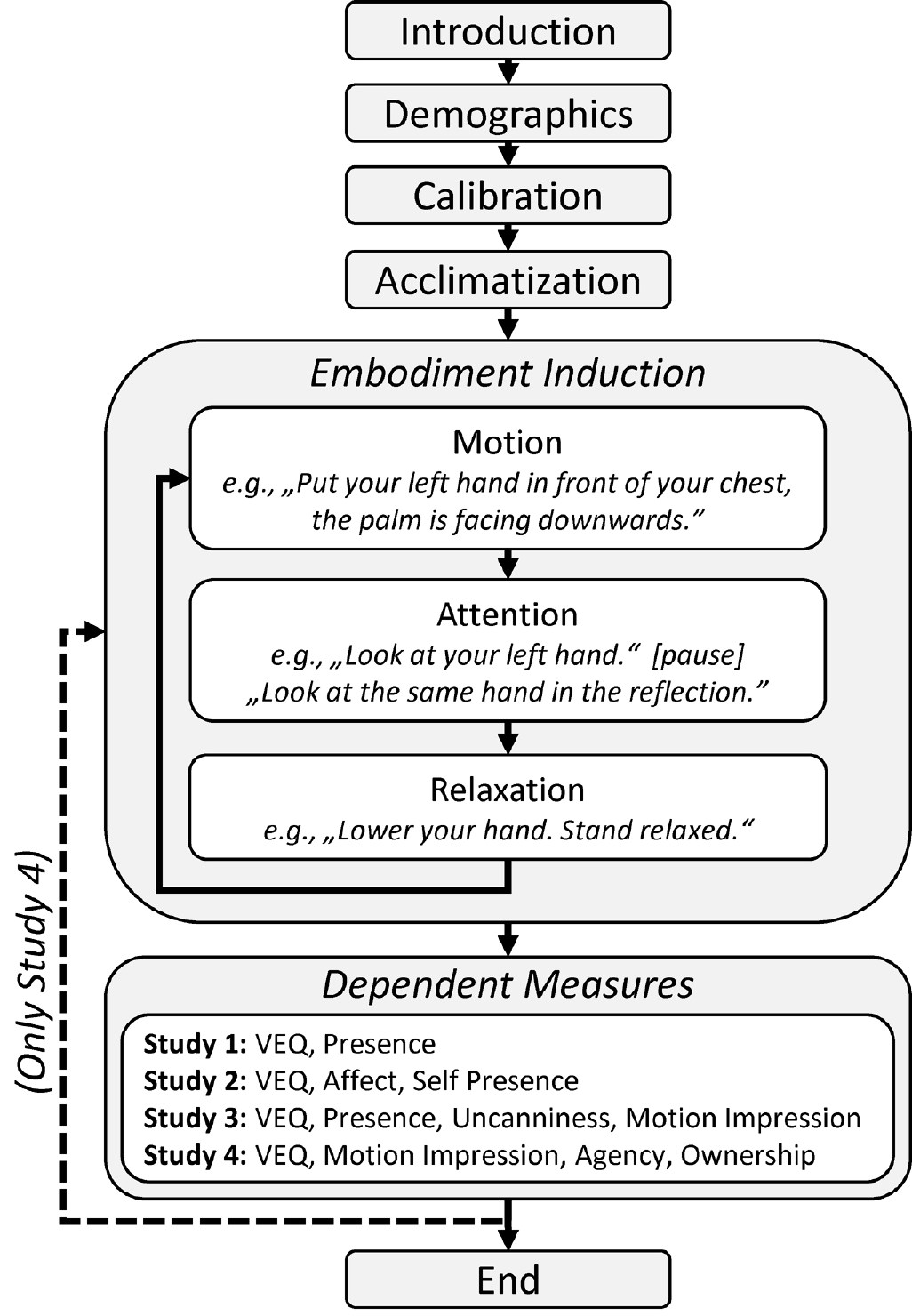} 
	\vspace{-2ex}
	\caption[General Study procedure.]{Study procedure.}
	\vspace{-4ex}
	\label{fig:procedure}
\end{figure}

\paragraph{Procedure}
\label{sec:genprocedure}
Fig. \ref{fig:procedure} shows the study procedure. Participants were welcomed and informed, before the pre-study questionnaire with demographics and media usage was assessed. Participants were then equipped, calibrated, and given time to acclimatize to the simulation. Similar to previous work \cite{waltemate_impact_2018,latoschik_fakemi_2016}, audio instructions were then presented to the users to induce embodiment with a total duration of 150 s. In these instructions, users were asked to perform actions, meaning to move certain body parts (e.g. ``Hold your left arm straight out with your palm facing down.''), and further focus on one's own/the avatar body parts, as well as the body parts of their avatar presented in a virtual mirror. (``Look at your left hand.'' [Pause] ``Look at the same hand in the reflection''), followed by a relaxing pose (``Put your arm comfortably back down and look at your reflection''). The full instructions are described in the supplementary material. Following the exposition, the dependent measures were assessed. 

\paragraph{Apparatus}
\label{sec:genapparatus}
The scenarios and materials are depicted in Fig. \ref{fig:immersion-apparatus}. Participants were tracked by an OptiTrack Flex 3 tracking system. A Unity3D simulation displayed the stimulus via a fake mirror projection or an Oculus Rift CV1 HMD (2160 px $\times$ 1200 px, 90 Hz, 120 degrees diagonal field of view). For the projection, the image was rendered using a fish tank VR \cite{Ware1993} approach with off-axis stereoscopic projection \cite{bourke_calculating_1999}, and displayed by an Acer H6517ST projector (projection size: sized 1.31 m high $\times$ 1.46 m width, active stero, 480 px $\times$ 1080 px per eye). The virtual projection was calibrated to match the physical projection preferences, thus allowing a physically accurate mirror image. The virtual camera/tracking point was constrained to the avatar head joint while accounting for the distances of the eyes (head-neck model). The baseline motion-to-photon latency was approximated by video measurement and frame counting (\cite{he_video-based_2000,waltemate_realizing_2015}) to 77 ms for the projection setup, assuming slightly lower values for the HMD setup. 
In the virtual environment, the participant was placed in a living room. A reference point was presented, as well as a virtual mirror (in the projection condition, the projection was the mirror; see Fig. \ref{fig:immersion-apparatus}). We used avatars created with Autodesk Character Generator (Fig. \ref{fig:immersion-apparatus}) which were scaled (uniform) according to the participant's height.

\paragraph{Measures}
We assessed the VEQ (Cronbach's $\alpha$ $\geq$.751) and the igroup presence questionnaire (IPQ) \cite{schubert_sense_2003-1}, which was adapted to fit the presented scenario.  
The IPQ adaptation assessed \textit{general presence} (``In the computer-generated world, I had a sense of `being there'''), \textit{spatial presence} (e.g., `` I did not feel present in the virtual space''; $\alpha = .786$), \textit{involvement} (e.g., ``I was not aware of my real environment''; $\alpha = .851$), and \textit{realness} (e.g., ``How real did the virtual world seem to you?''; $\alpha = .672$). The responses were given using a 7-point Likert-type scale (see the original source for the anchors). Further assessments are not the subject of the present reporting due to page limitations. No severe sickness effects occured.

\paragraph{Participants}
We excluded participants when problems or severe tracking errors were noted.
The final sample consisted of 50 participants ($32$ female, $18$ male, $M_{age}=22.18$, $SD_{age}=2.83$). Of those, $49$ participants were students, and $46$ participants had previous experience with VR technologies. The sample was equally distributed (25 per condition).

\subsubsection{Results}

\paragraph{Comparisons}

T-tests were conducted for each individual measure. The VEQ factors were aggregated according to the scoring depicted in Table \ref{tab:goldivbo}. In the case of unequal variances (Levene-test), corrected values are reported (Welch-test). 
The perceived \textit{change} in body schema was statistically significantly different between the projection condition ($M=2.61$, $SD=1.46$) and the HMD condition ($M=3.40$, $SD=1.31$; $t(48)=-2.013$, $p=.0498$, $d=0.61$). As expected, the HMD condition resulted in a stronger perception of perceived change. Neither ownership (projection: $M=4.81$, $SD=0.85$; HMD: $M=4.71$, $SD=1.22$) nor agency (projection: $M=6.30$, $SD=0.51$; HMD: $M=6.18$, $SD=0.72$) showed a significant difference ($p$s$\ge.497$). 
Regarding the IPQ, we found a significant difference with general presence between the projection condition ($M=3.88$, $SD=1.45$) and the HMD condition ($M=5.32$, $SD=0.99$); $t(42.30)=-4.098$, $p<.001$, $d=1.16$.
This effect was substantiated by a significant difference in the spatial presence measure between the projection condition ($M=3.88$, $SD=1.32$) and the HMD condition ($M=5.03$, $SD=1.20$); $t(48)=-3.218$, $p=.002$, $d=1.19$, as well as for the involvement measure between the projection condition ($M=3.45$, $SD=1.28$) and the HMD condition ($M=5.44$, $SD=1.18$); $t(48)=-5.712$, $p<.001$, $d=1.62$. There was no significant difference in the realness measure $t(48)=-0.388$, $p=.70$.

\paragraph{Correlations}	
We calculated bivariate Pearson correlations. Table \ref{tab:immersion:corrs} depicts the results. 
We found significant correlations between \textit{ownership} and \textit{agency}, as well as between ownership and general presence, spatial presence, and realness. The \textit{change} factor was correlated significantly with the general presence assessment. Further, we found correlations within the presence measures.

\begin{table}
	\caption[Pearson correlations (Study 1).]{Significant Bivariate Pearson Correlations(r) -- Study 1.}
	\label{tab:immersion:corrs}
	\scriptsize%
	\centering%
	\vspace{-1ex}
	\begin{tabu}{%
			*{7}{l}%
		}			 
		\midrule
		& AG 	& CH & GP & SP & IN & RE  \\
		\midrule
		Ownership 			& $.46$** 	&  		 & $.36$*			& $.34$*		   &  		& $.42$**      \\
		Change 				& 			&  		 & 	$.30$*			&  		   			&  		 		&       \\
		GP 	& 			&  		 & 					&  $.75$**  		&  	$.59$** 	&   $.53$**    \\
		SP 	& 			&  		 & 					&  		   			&  	$.63$** 	 &   $.40$**  \\
		\midrule
		\multicolumn{7}{l}{\em{Note. *$p<.05$; **$p<.01$. AG Agency, CH Change, GP General}}\\
		\multicolumn{7}{l}{\em{Presence, SP Spatial Presence, IN Involvement, RE Realness.}} \vspace{-5ex} \\	
	\end{tabu}
\end{table}

\begin{figure*}[t]
	\centering
	\frame{\includegraphics[height=1.3in]{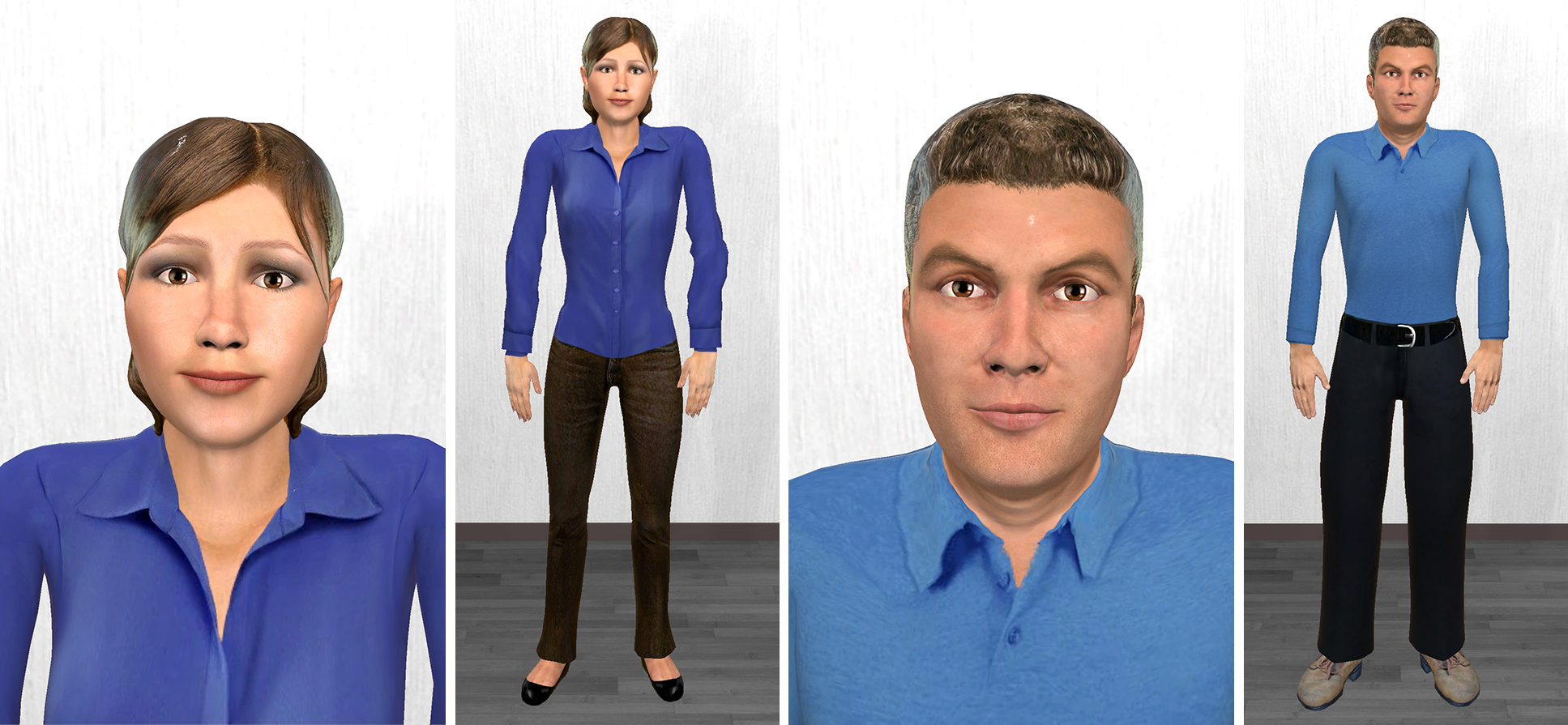}}\hfill	\frame{\includegraphics[height=1.3in]{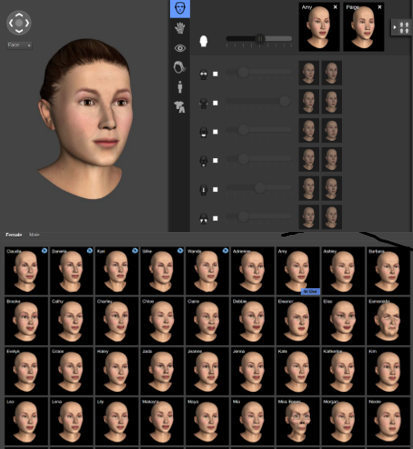}}\hfill
	\frame{\includegraphics[height=1.3in]{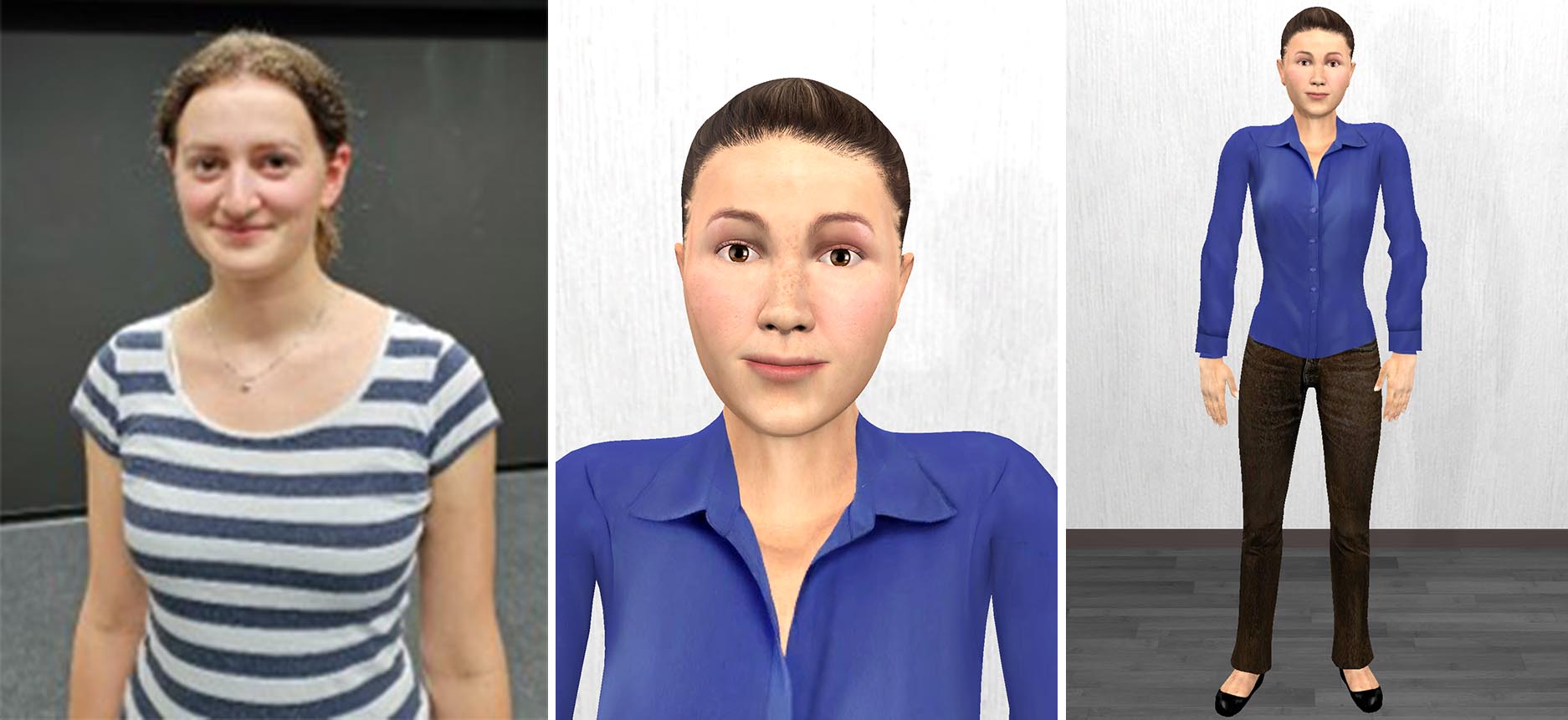}} 
	\vspace{-2.5ex}
	\caption[Character examples (Study 2).]{\textbf{Apparatus.} \textit{Left:} Generic characters (Study 2, 3, and 4).  \textit{Center:} Character generator.  \textit{Right:} Personalization example (Study 2).} \vspace{-3ex}
	\label{fig:chargen}
\end{figure*}

\subsubsection{Discussion}
The significantly higher perception in the \textit{change} of the perceived body schema substantiates previous findings that showed that higher immersion positively impacts embodiment (H1.1). While Waltemate, et al. \cite{waltemate_impact_2018} also found immersion to impact ownership over a virtual body, the proposed scale picked up that the first-person perspective HMD simulation specifically affected the perceived change of one's own body schema. We interpret this as a result of the presence of a reference to the physical body in the projection condition, and the absence of such in the HMD condition. This indicates a more sensitive pinpointing of this effect by the VEQ. This was confirmed by the expected correlations between \textit{ownership} and \textit{agency} as suggested in previous literature, and an absence of correlations of these components with \textit{change}. Despite relatively high ratings for agency, and above scale mean ratings for ownership, these factors were not significantly impacted between conditions.

Supporting H1.2, several presence dimensions were affected. The HMD condition resulted in greater presence. In addition, we found positive relations between the VEQ and presence measures. \textit{Ownership} correlated with general presence, spatial presence, and realness, whereas the change factor correlated only with the general presence assessment. Therefore, we assume that presence and embodiment interact positively. However, we cannot provide any insights into the causality based on the analysis. 
We concluded that immersion is a driving factor for embodiment regarding the fostering of the \textit{change} in the perceived body schema, as more immersive HMD-driven simulations allow for the perception of a virtual body from a first-person perspective without a physical body reference.

\subsection{Study 2: User-Performed Personalization}
\label{sec:beta-personalization}

Previous research investigated the use of photogrammetric avatars and found that personalization can affect the identity with \cite{lucas2016avatars}, and the ownership over, a virtual body \cite{jung2018over,waltemate_impact_2018,latoschik_fakemi_2016}. Based on these results, we hypothesized H2.1: \textit{Personalization increases the perception of ownership over a virtual body}. However, whether this assumption holds true using avatar creation tools instead of photogrammetric scanning is an open question and the topic of Study 2.

\subsubsection{Method}
In a one-factor (\textit{Personalization}) between-subjects design, we compared the embodiment with an avatar from a creation tool to a personalized avatar created with the same tool. Participants were represented either as a gender-matched generic avatar (Caucasian--as we expected a Caucasian sample) or by a personalized avatar they created as a representation of themselves (see Fig. \ref{fig:chargen}). 

\paragraph{Procedure}
The procedure followed the general study procedure (Fig. \ref{fig:procedure}). Participants were welcomed and informed, before the pre-study questionnaire was assessed. Participants were then asked to create an avatar (personalized condition) or to inspect the generic avatar (generic avatar condition). In the personalized condition, participants were taught the avatar generator software and provided help. Participants were permitted to modify the virtual character's body measures (form, proportions), facial appearance, and hairstyle in 15 min of preparation time. The clothing was kept similar in the two conditions. Following the avatar creation/inspection, a similarity measure was assessed. Participants were then calibrated and had time to acclimatize to the simulation. Similar to Study 1, audio instructions asking participants to perform movements and focus on body parts were used for the embodiment exposition. In addition to the instructions for Study 1, the participants were specifically instructed to step closer to the mirror and pay attention to features of the character's appearance (e.g., ``Look at the eyes of the mirrored self,'' ``Look at the mouth of the mirrored self,'' ``Turn 90 degrees and look at the mirrored self from the side''). The instructions lasted for 180 s and are described in detail in the supplementary material. 

\paragraph{Apparatus}
The apparatus consisted of a setup identical to the HMD-based setup of Study 1, except that a FOVE 0 HMD (2560 px $\times$ 1440 px, 70 Hz, 100 degrees field of view) was used as the display.

\paragraph{Measures}
\label{sec:goldivbo-personalization-measures}
We measured demographic variables, the VEQ ($\alpha$s $\geq .744$), and affect using the PANAS scale in the short form \cite{thompson_development_2007} (PA: $\alpha= .840$; NA $\alpha = .319$, dropped from analyses). As we expected that personalization may also have an impact on self-presence, we assessed Ratan and Hasler's self-presence questionnaire \cite{ratan_exploring_2010}, adapted to the context of the study (excluded: ``To what extent does your avatar’s profile info represent some aspect of your
personal identity?'' and ``To what extent does your avatar’s name represent some aspect of your
personal identity?'').
Proto--self-presence ($\alpha =.781$) was assessed with items such as ``How much do you feel like your avatar is an extension of your body within the virtual environment?''
Core self-presence ($\alpha =.838$) was assessed with items such as ``When arousing events would happen to your avatar, to what extent do you feel
aroused?''
Extended self-presence ($\alpha =.661$) was assessed with items like ``To what extent is your avatar’s gender related to some aspect of your personal
identity?'' (5-point scale, see \cite{ratan_exploring_2010}).
To control for the manipulation, we measured the perceived similarity toward the avatar (generic or personalized) before the exposure (desktop monitor), and after the exposure (reflecting the experience): ``Please rate how much the virtual character is similar to you on the following scale, where 1 equals no similarity and 11 equals a digital twin.'' Further measures such as sickness and humanness are not part of the present discussion. No severe sickness effects occurred.

\paragraph{Participants}
We excluded participants when problems or severe tracking errors were noted.
The final sample consisted of participants (generic avatar: $N=25$, personalized avatar: $N=23$. $27$ female, $21$ male, $M_{age}=21.64$, $SD_{age}=2.25$). All $48$ participants were students, and $45$ participants had previous experience with VR. 

\subsubsection{Results}

\paragraph{Comparisons}
T-tests did not reveal significant differences for the VEQ factors. The mean values for ownership were higher in the personalized condition ($M=4.85$, $SD=0.84$) compared to the generic condition ($M=4.36$, $SD=1.26$), but not to a significant level ($p=.124$). Similar images appeared regarding agency (personalized: $M=6.11$, $SD=0.56$; generic: $M=5.84$, $SD=0.79$), and change (personalized: $M=3.55$, $SD=1.53$; generic: $M=2.97$, $SD=1.30$); $p$s$\geq .159$.
We found a significantly higher positive affect for the personalized avatar condition ($M=3.64$, $SD=0.68$) compared to the generic avatar condition ($M=3.14$, $SD=0.64$; $t(46)=2.655$, $p=.011$, $d=0.75$). 
We found a significant effect for proto-self-presence, showing higher ratings for the personalized avatar condition ($M=3.59$, $SD=0.67$) compared to the generic avatar condition ($M=3.04$, $SD=0.81$; $t(46)=2.528$, $p=.015$, $d=0.734$). 
The perceived pre-exposure similarity with the virtual character was higher for the personalized condition ($M=5.30$, $SD=1.72$) compared to the generic condition ($M=4.64$, $SD=1.98$), but not to a significant level. In the post-exposure measurement, these differences were almost equal (personalized: $M=5.13$, $SD=1.71$; generic: $M=5.16$, $SD=2.04$).


\paragraph{Correlations}
Bivariate Pearson correlations are presented in Table \ref{tab:personalization-goldivbo-correlations}.
The \textit{ownership} factor correlated with \textit{agency} and \textit{change}. \textit{Agency} and \textit{ownership} correlated with the post-exposure similarity measure, whereas the \textit{change} factor did not. All embodiment factors correlated with the proto-self-presence measure. Interestingly, the \textit{agency} factor correlated with a perceived positive affect.

\begin{table}[tb]
	\caption[Pearson correlations of related constructs (Study 2).]{Significant Bivariate Pearson Correlations ($r$) -- Study 2.}
	\label{tab:personalization-goldivbo-correlations}
	\scriptsize%
	\centering%
	\vspace{-1ex}
	\begin{tabu}{%
			*{11}{l}%
		}
		
		\midrule
		&   AG & CH & PSP  & CSP & PA    & SPre & SPo    \\
		
		
		\midrule
		
		OW    & .51** & .37** & .70**  &    &  &  & .33*   \\ 
		
		
		
		AG    &  &  & .65**  & .35* & .43**   &  & .31*   \\ 
		
		
		
		CH  &  &  & .30*  &  &  &  &  &    \\ 
		
		
		
		PSP   &  &  &   & .37* & .33* &   & .43**   \\ 
		
		
		
		ESP    &  &  &   & .40** &  &   .36* & .32*   \\ 
		
		
		
		CSP    &  &  &   &  & .31* &    & .43**   \\ 
		
		
		
		PA    &  &  &   &  &  &  .35* & .31*   \\ 
		
		
		
		
		
		
		SPre    &  &  &   &  &  &  &   .44**   \\ 
		
		\midrule
		\vspace{-2ex}\\
		\multicolumn{11}{l}{\em{Note. * $p<.05$;  ** $p<.01$. OW ownership, AG agency, CH change,}}\\
		\multicolumn{11}{l}{\em{PSP proto--self-presence, CSP core self-presence, PA positive affect,}}\\ 
		\multicolumn{11}{l}{\em{SPre similarity pre-exposure, SPo similarity post-exposure.}}\\
		\vspace{-5ex}
	\end{tabu}%
\end{table}

\subsubsection{Discussion}
Contrasting our assumption and previous work on photogrammetric personalization \cite{latoschik_fakemi_2016,waltemate_impact_2018}, we did not find evidence for H2.1. The personalization procedure did not significantly affect ownership, or other VEQ factors. A mere self-performed personalization did not provoke ownership to a strong degree, which is partially also supported by the post-exposure similarity measure. The results are limited by the fact that the participants' clothing was not personalized and that the creation was limited in time, as well as the participants' software skillset. Yet, the manipulation affected the perception, as participants perceived higher self-presence in the simulation. 
We found a positive association between the post-exposure similarity, ownership and agency, confirming their relation. 
The correlations between the VEQ factors and the self-presence measures point at a relationship between embodiment and the perception presence, potentially feelings and impressions that may be evoked by simulation events when controlling an avatar.

We concluded that the self-performed, character generator-based personalization procedure did not evoke significanly higher perceived embodiment, contrasting with the results for photogrammetric personalization. Further, the study showed that covariance is shared between aspects of self-presence and embodiment.

\subsection{Study 3: Behavioral Realism}
\label{sec:beta-realism}
A comparison of different degrees of behavioral realism and the impact on virtual embodiment, to our knowledge, has not yet been investigated. As agency specifically relates to controlling one's own movements \cite{tsakiris_having_2006}, we hypothesized H3.1: \textit{Increased behavioral realism increases the perceived agency over a virtual body}. 

\subsubsection{Method}
In a two-factor (\textit{Facial Expressions}, \textit{Gaze}) between-subjects design, we evaluated the impact of behavioral realism. Participants were represented by generic avatars (see Fig. \ref{fig:chargen}, top) and randomly assigned to one of four conditions: body motion only ($BO$), body and facial motion replication ($BF$), body and gaze motion replication ($BG$), and body, gaze, and face motion replication ($BFG$).

\paragraph{Procedure}
The procedure followed the general procedure (Fig. \ref{fig:procedure}).
Participants were welcomed and informed, before we assessed the pre-study questionnaire. Participants were then calibrated and had time to acclimatize. Audio instructions then asked participants to perform bodily movements and focus on body parts. In Study 3, the participants were specifically instructed to move closer to another marking in front of the mirror, and let their gaze wander to specific focus points, trying to ensure an influence in perception of the manipulation (e.g., ``Fixate on the left eye of your mirrored-self,'' ``Focus on the right ear of your mirrored self'' etc.). For the facial expressions, we asked participants to perform certain expressions (e.g., ``Open and close your mouth,'' ``Try to express happiness by smiling at your mirrored-self,'' etc.). The instructions lasted for 219 s and are described in the supplementary material. After the exposure, we assessed the dependent measures.

\begin{figure}[t]
	\centering
	\includegraphics[width=1\columnwidth]{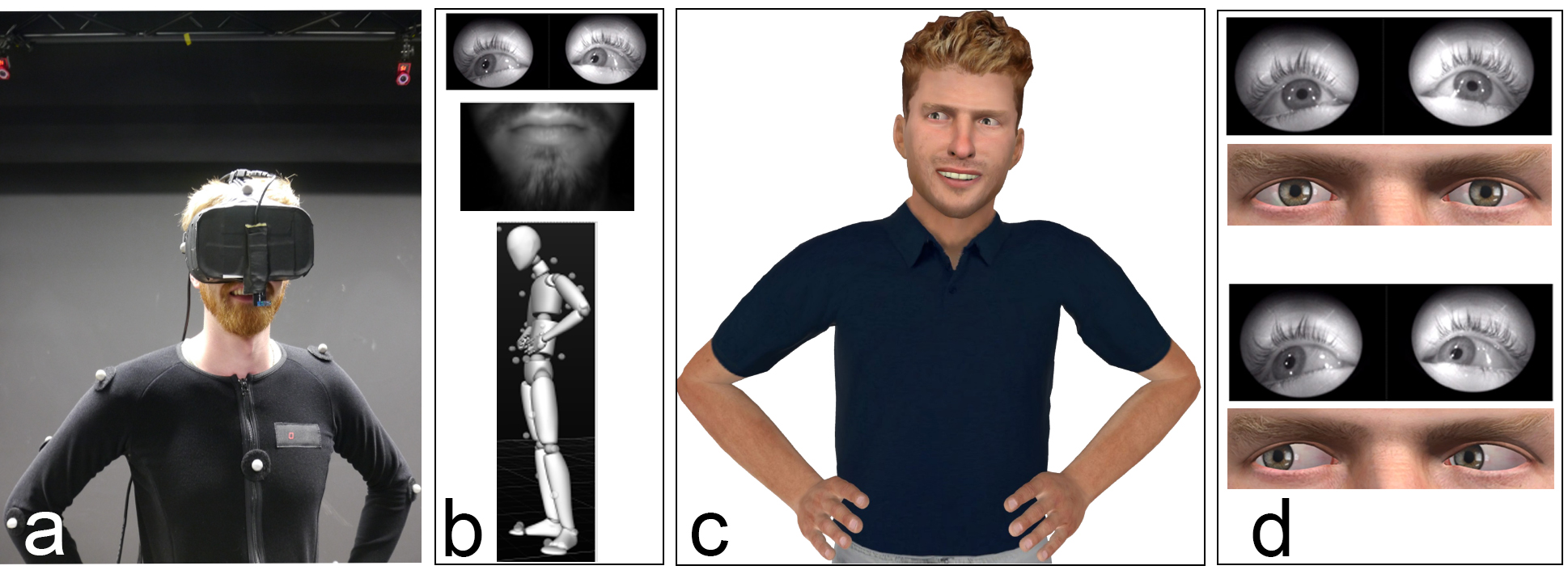}\\ 
	\vspace{-2ex}
	\caption[]{\textbf{Sensing and replication (Study 3).} \textit{a)} Tracking. \textit{b)} Sensory data (exemple images). \textit{c} Replication. \textit{d} Gaze detail example. The avatars that were used in the study are depicted in Fig. \ref{fig:chargen}.} \vspace{-2ex}
	\label{fig:behavioralrealism}
\end{figure}

\paragraph{Apparatus}
We used the same apparatus and generic avatars as in Study 2, and included the tracking of the participant's gaze using the FOVE 0's eye tracking system as well as facial expression tracking performed by a BinaryVR Dev Kit V1. 
Combining the two individual eye vectors, the FOVE integration calculates the intersection point to estimate the convergence point in the virtual scene, which is the approximate focus point of the user \cite{wilson_head_2017}. From this position, we recalculated the eyeball rotation. The BinaryVR Dev Kit was used to gather information about lower facial deformation. The 3D depth-sensing device Pico Flexx (up to 45 Hz) was affixed to the HMDs. 
Its 2D and depth images were processed by the BinaryVR Dev Kit to generate facial deformation parameters \cite{jihun_head-mounted_2017,yu_real-time_2016}. Tracked expression parameters were mapped to blendshapes, for example, jaw open, and smile.
The body tracking, facial expression, and gaze data was fused in the simulation to drive the avatar (see Fig. \ref{fig:behavioralrealism}).

\paragraph{Measures}
We assessed the VEQ ($\alpha$s $\geq .732$), a rating of the avatar assessing humanness, eeriness, and attractiveness ($\alpha$s $\geq .688$) \cite{ho_revisiting_2010}, the self-presence measures previously applied in Study 2 ($\alpha$s $\geq .645$) \cite{ratan_exploring_2010}, as well as affect ($\alpha$s $\geq .674$) \cite{thompson_development_2007,janke_deutsche_2014}. We further asked how real, how natural, and how synchronous the motion behavior of the avatar appeared to the participants: ``The movements were realistic,'' ``The movements were naturalistic,'' ``The movements were in synchrony to my own movements'' (1--strongly disagree, 7--strongly agree). Further measures were excluded due to page limitations. No severe sickness effects occurred.

\paragraph{Participants}
We excluded participants when problems or severe tracking errors were noted. 
The final sample for the analysis consisted of 70 participants ($46$ female, $24$ male, $M_{age}=21.3$, $SD_{age}=1.82$, all students), of whom $65$ had previous VR experience. 17 participants were assigned to the $BO$ condition, 18 to $BF$, 18 to $BG$, and 17 to $BFG$.  

\begin{figure*}[t]
	\centering
	\frame{\includegraphics[height=1.03in]{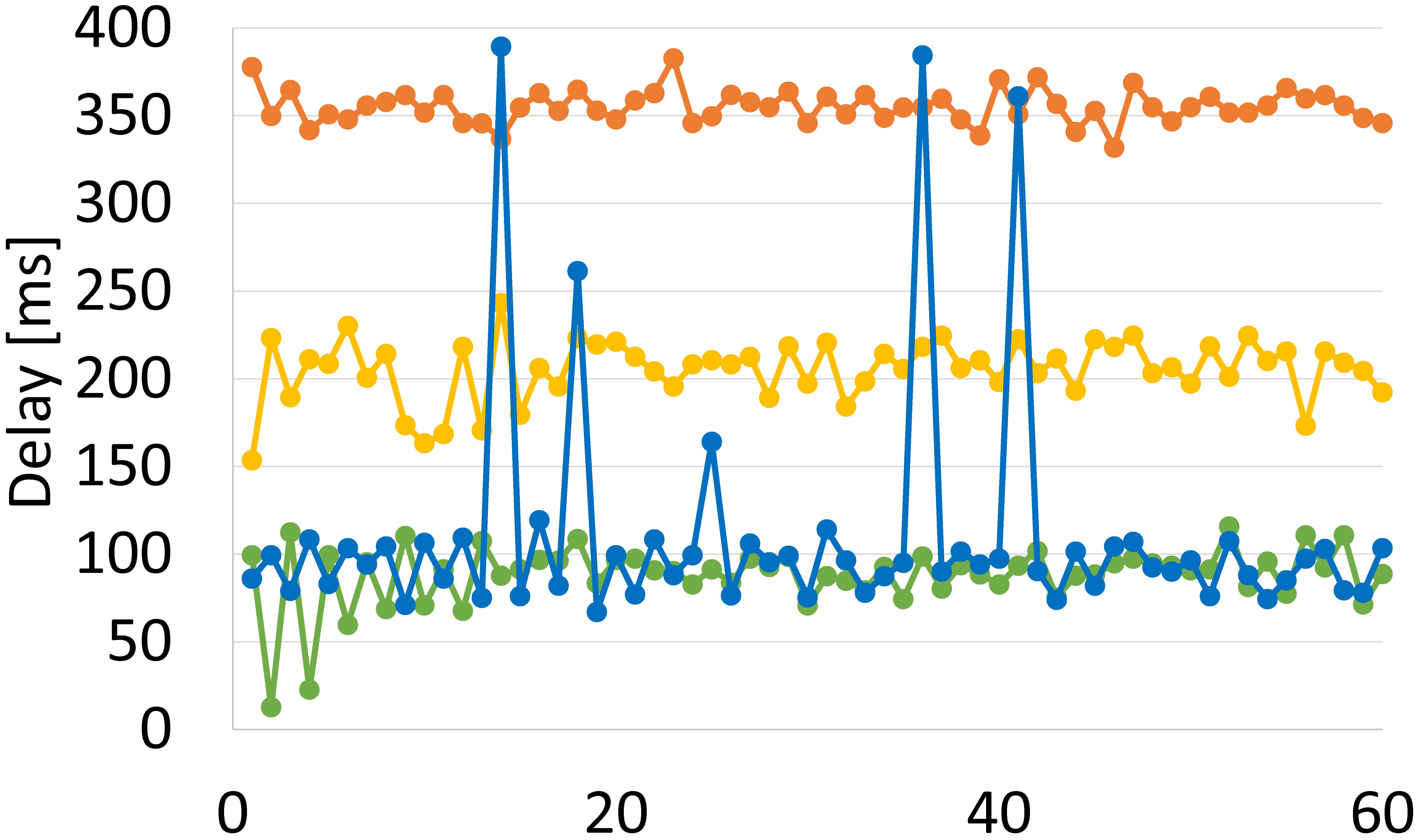}} \hfill 
	\frame{\includegraphics[height=1.03in]{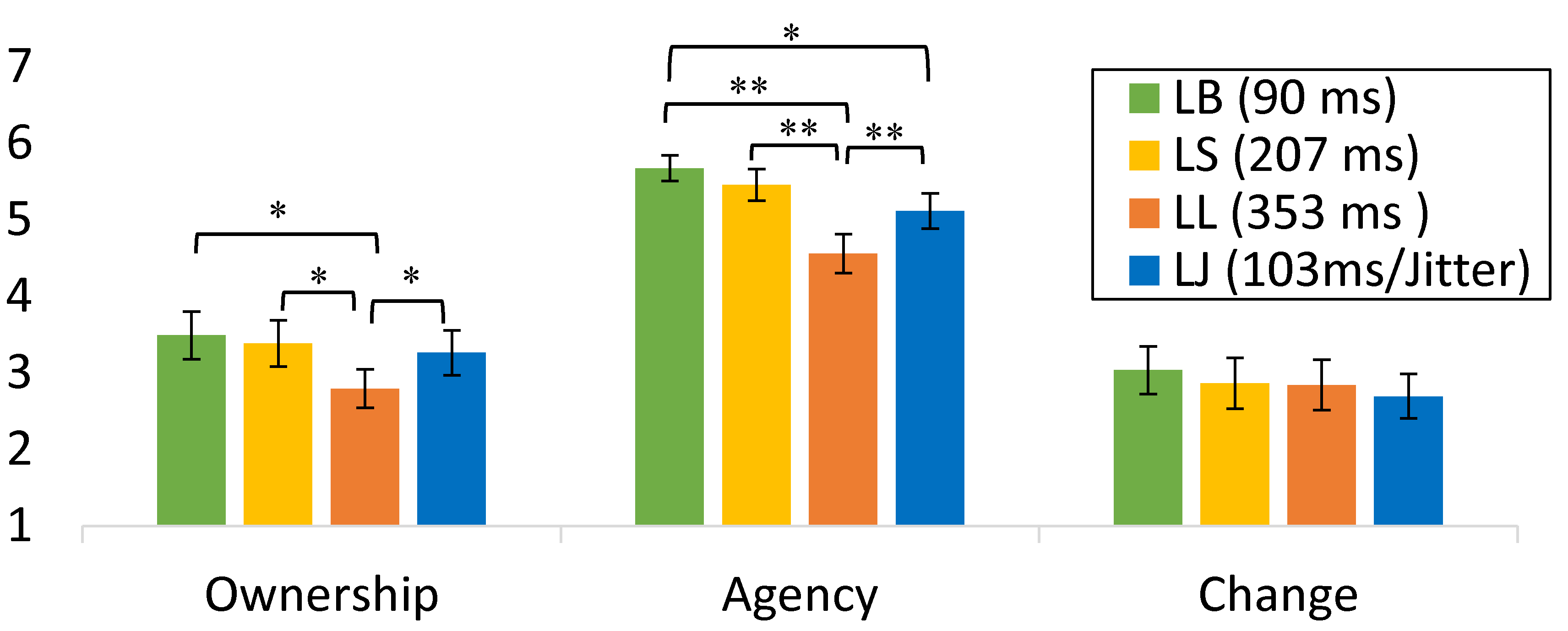}} \hfill
	\frame{\includegraphics[height=1.03in]{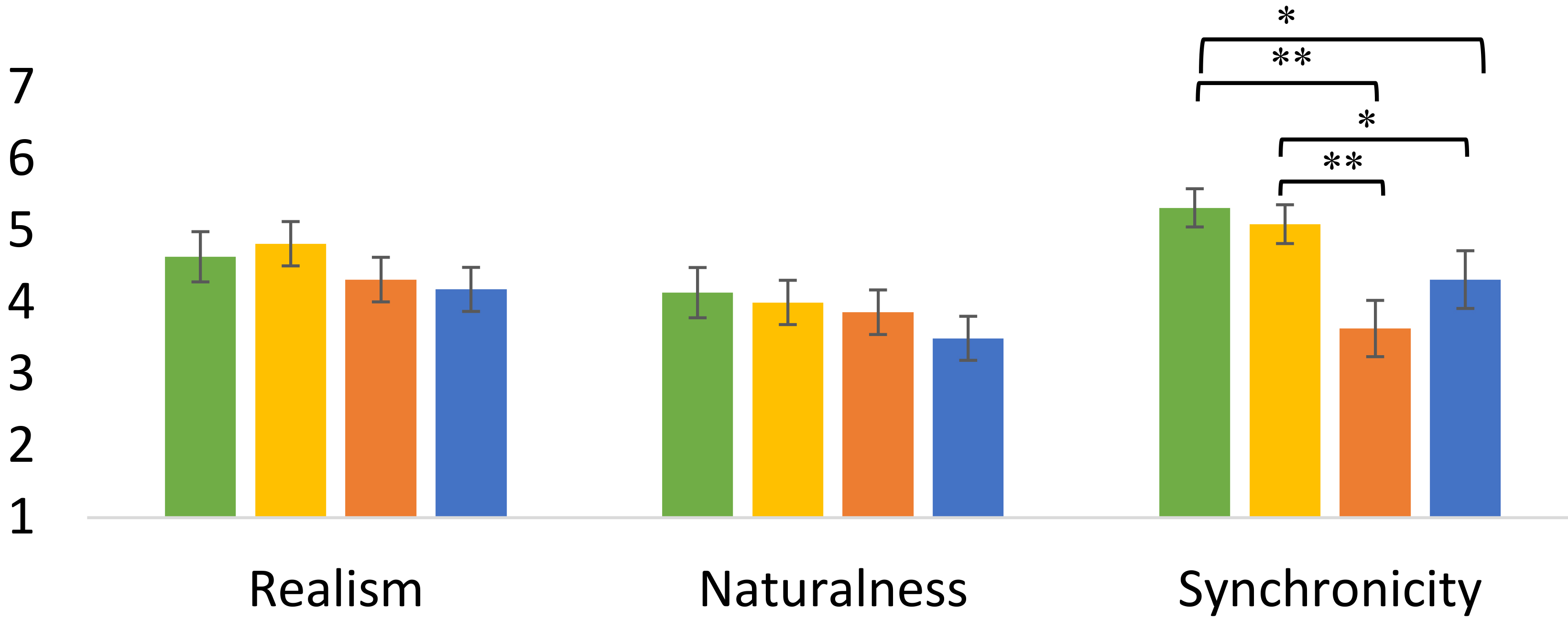}}
	\vspace{-4ex}
	\caption[Latency manipulations]{\textbf{Latency manipulation and resulting impacts.} \textit{Top:} 60 sample measures of the induced latency assessed by frame counting. {Bottom: Descriptive results of the VEQ assessment. \textit{Note. Error bars denote standard errors. *** $p<.001$; ** $p<.01$; *$p<.05$.}}} \vspace{-3ex}
	\label{fig:latency}
\end{figure*}

\subsubsection{Results}
\paragraph{Comparisons}
We calculated two-factor (gaze, facial expression) analyses of variance (ANOVAs).
Although the $BFG$ condition was rated highest in ownership ($M$s between $4.28$ and $4.68$) and highest in control ($M$s between $5.65$ and $5.99$), the differences were not significant. We found a significant effect of gaze on the perceived humanness; $F(1,66)=7.826$, $p=.007$, $\eta_p^2=.106$, indicating that the perceived humanness was higher in the conditions with enabled gaze tracking and replication ($M=3.07$, $SD=0.78$) compared to the conditions with disabled gaze tracking and replication ($M=2.60$, $SD=0.60$). No further significant effects were observed.

\subsubsection{Discussion}
We did not find supporting evidence for H1.3. Increased behavior realsim did not significantly impact the perceived embodiment between conditions. The additional gaze replication resulted in a greater perception of humanness. Regarding the perceived agency, we interpret that the facial expression and gaze replication did not have a strong influence in comparison to body movement, which was always present and represents greater motion dynamics, and arguably evokes a stronger visual stimulation. Future research should investigate body movement as an additional manipulation factor. Some limitations arise. The detection of facial and gaze behavior was limited by the visual resolution of the display. However, the participants stood especially close (about 50 cm) to the virtual mirror during parts of the induction phase. Third, the induction phase might have been too short for some participants, and as a result, the time spent on the gaze and facial expression interaction might not have been sufficient. 

\subsection{Study 4: Latency and Latency Jitter}
\label{sec:intra:latency}
In an independent fourth study, we further validated the scale, specifically targeting the agency factor. Previous work suggested that latency negatively impacts virtual embodiment \cite{waltemate_impact_2016}. Thus, we hypothesized H4.1: \textit{Latency negatively impacts components of virtual embodiment}. Although linear latencies were subject to previous investigations \cite{waltemate_impact_2016}, the impact of latency jitter has not been assessed.


\subsubsection{Method}
In a one-factor (\textit{Latency Level}) repeated-measures design, we evaluated the impact of latency and latency jitter, meaning the non-periodic spontaneous peaks of latency, on the perception of the factors of the VEQ. 
Participants were exposed to four conditions of delayed and jitter-delayed simulation display (baseline, LB; small latency, LS; large latency, LL; latency jitter, LJ).

\paragraph{Procedure}
The procedure followed the procedure depicted in Fig. \ref{fig:procedure} in repeated fashion. We welcomed participants and informed them, before assessing the pre-study questionnaire.
The exposure conditions were then presented to the participants in randomized order. In each trial, the participants were calibrated, exposed to the simulation and induction, followed by an assessment of the dependent measures.
In the audio instructions, the participants were specifically instructed to perform more rapid and fluid movements (e.g., ``Raise your left arm at moderate speed in front of you, and lower it back down next to your hip. Repeat this movement ten times, and focus on your arm while doing so''). The complete audio instructions are described in the supplementary material. The instructions lasted for 282 s. 

\paragraph{Apparatus}
We used a similar apparatus Study 2. By buffering the tracking input data from the motion tracking system, artificial delays were introduced into the simulation by buffering and delaying the replication of the body motion tracking data. We prevented biasing sickness effects and \textbf{influenced only the delay of the body movements, but did not influence the delay of the virtual camera (head movements, respectively)}, which, therefore, transformed according to the raw system delay without further modifications. Thus, the body motion was delayed, whereas the camera/head pose was not. We adapted the procedure described by Stauffert, et al. \cite{stauffert_effects_2018} to introduce latency jitter, which uses a stochastical model for latency distribution, introducing high-latency spikes into the simulation.
Motion-to-photon latency of the resulting simulations was approximated by video frame counting (Canon, 1000 Hz). 
The measures resulted in $M=90.12$ ms ($SD=16.14$ ms) for the simulation baseline $L_B$, $M=206.93$ ms ($SD=16.47$ ms) for the small delay $L_S$, $M=353.07$ ms ($SD=15.38$ ms) for the larger delay $L_L$, and  $M=102.58$ ms ($SD=49.71$ ms) for $L_J$. $L_B$, $L_S$, and $L_L$ were measured with 60 samples (see Fig. \ref{fig:latency}). Despite measuring $N=165$ repetitions in the jitter condition $L_J$, the resulting mean and SD may not accurately reflect the induced jitter, due to some spikes that could not be captured using the motion-apex measurement applied. Users were embodied with the generic avatars (see Fig. \ref{fig:chargen}).


\paragraph{Measures}
We assessed the VEQ ($\alpha$s $\geq .771$) and performed a comparison to the questionnaire developed for RHI experiments by Kalckert and Ehrsson (KE) \cite{kalckert_moving_2012} (partly adapted from \cite{longo_what_2008}), that measures ownership ($\alpha$s $\geq .801$), ownership control ($\alpha$s $\geq .655$), agency ($\alpha$s $\geq .636$), and agency control ($\alpha$s between $.239$ and $.563$) with a 7-point Likert-type scale (1--strongly disagree, 7--strongly agree). Questioning was adapted to fit the scenario (e.g., ``The rubber hand moved just like I wanted it to, as if it was obeying my will'' = ``The virtual body moved just like I wanted it to, as if it was obeying my will'').
As manipulation control, we asked how realistic, natural, and synchronous the movements of the avatar appeared to the participants: ``The movements were realistic,'' ``The movements were naturalistic,'' ``The movements were in synchrony to my own movements'' (1--strongly disagree, 7--strongly agree).
Further measures are not part of the present discussion. No severe sickness effects occurred.

\paragraph{Participants}
We excluded participants when problems or severe tracking errors occured. The final sample consisted of 22 participants ($17$ female, $5$ male, $M_{age}=21.77$, $SD_{age}=3.62$, $22$ students), of whom $21$ had previous experience with VR. 

\subsubsection{Results}
\paragraph{Comparisons}
We calculated repeated-measures ANOVAs for each dependent variable. Where the assumption of sphericity was violated, we report Greenhouse--Geisser corrected values. Fig. \ref{fig:latency} depicts the descriptive results.
We found a significant main effect for \textit{ownership}; $F(3,63)=3.57$, $p=.024$, $\eta_{p}^2=.138$. Similarly, \textit{agency} measure was affected; $F(1.741,63.553)=7.50$, $p=.003$, $\eta_{p}^2=.263$. No significant impacts were observed for \textit{change}. We did not observe any significant main effects in the agency and ownership measures of the scale adapted from KE \cite{kalckert_moving_2012}. The synchronicity assessment showed a significant main effect $F(1.905,40.005)=7.077$, $p=.003$, $\eta_{p}^2=.252$.
In contrast, neither the realism, nor the naturalness of behavior showed a significant main effect. 
The strongest linear latency injection (LL) yielded to the lowest scoring of ownership and agency (see Fig. \ref{fig:latency}). The jitter condition was similarly affected, but resulted in significantly better ratings than the condition with the largest latency injection. The lower level linear latency (LS) had comparable results to the jitter-injected condition. Similarly, the synchronicity ratings were affected (see Fig. \ref{fig:latency}). 



\paragraph{Correlations}
Pearson correlations are depicted in Table \ref{tab:latency-goldivbo-correlations}.
The synchronicity assessment showed large correlations with the \textit{agency} factor of the VEQ and the agency factor of the scale from KE \cite{kalckert_moving_2012}. Furthermore, synchronicity showed a medium to large correlation with the \textit{ownership} factor, and the KE ownership factor. 
In turn, \textit{ownership} showed medium to large correlations with \textit{agency}, and a large correlation with ownership (KE) and ownership control (KE) across all measures. 
Interestingly, we did not find stable correlations between the perceived naturalness, synchronicity, and realism of the movement, providing room for further discussion.

\begin{table}[tb]
	\caption[Pearson correlations of the latency manipulation (Study 4).]{Significant Bivariate Pearson Correlations -- Study 4.}
	\label{tab:latency-goldivbo-correlations}
	\vspace{-1.5ex}
	\scriptsize%
	\centering%
	\begin{tabu}{%
			*{8}{l}%
		}			 
	
		\midrule
		 		  		& OW  	& AG  & CH  & AG & AC & OW  & OC      \\ \vspace{-.5ex}
		 		&   	&   &  &  KE &   KE &  KE &   KE       \\ 
		
		\midrule \vspace{-2ex}
		\\ 
		
		   \scriptsize   MR $L_B$     &   & .58**   &    & .71**   &     &        &        \\
		  MR $L_S$  & .48* & .50*	& & .67**
		\\
		 MR $L_L$  & & .49*
		\\
		 MR $L_J$  &	.53* &	.68**  & &	.68** & & .49* &
		\\ \vspace{-2ex}
		\\

		           MN  $L_B$                  & .45*              & .51*       &                     & .55**               \\
		  MN $L_S$  & .49*
		\\
		MN $L_L$  & & & & & & & .45*
		\\
		 MN $L_J$  &	.66**	& .46*	&&	.43* &&	.59**&	.60**
		\\ \vspace{-2ex}
		\\

	    MS $L_B$                          &                    & .57**      &                     & .80**         &                        &        &        \\
		  MS $L_S$  & .61** & .66** & & 	.54**	& &	.44* &	.50*  \\
		 MS $L_L$   &	.47*&	.80**&	&	.66**& &	.62**
		\\
		 MS $L_J$  	&	.59** &	.77**	&&	.55**	&&	.61** &	.45* 
		\\ \vspace{-2ex}
		&\\
		
		  OW $L_B$                        &                    & .48*       &                     &                & .52*                  & .84** & .71** \\
		  OW $L_S$ & & .53*	& & &	.50* &	.88** &	.72** 	\\
		 OW $L_L$ & & .44* & &		& &	.90**	 & .62**
		\\
		 OW $L_J$	&	&	.56**	&& 	&	&	.93** &	.67**
		\\ \vspace{-2ex}
		&\\
		
		       AG  $L_B$                        &                    &             &                     & .74**         &                        &        &        \\
		  AG $L_S$  & & & &	.71** & &	.47* &  \\
		 AG $L_L$ 	&	&	&	&	.83**	& &	.58** &
		\\
		 AG $L_J$ 	&	&	&	&	.79**	&&	.53*	&
		\\ \vspace{-2ex}
		\\
		
	  CH $L_S$  & & & & &	.48* \\
		
%
%
%
		\midrule 
		\multicolumn{8}{l}{\em{Note. * $p<.05$; ** $p<.01$; MR movement realism, MN movement natu-}}\\
		\multicolumn{8}{l}{\em{ralism, MS movement synchronicity, OW ownership, AG agency, CH }}\\
		\multicolumn{8}{l}{\em{change; Comparison measures: AGKE agency, ACKE agency control, }}\\
		\multicolumn{8}{l}{\em{OWKE ownership, OCKE ownership control, KE adapted from \cite{kalckert_moving_2012}.}}
		\vspace{-4ex}
	\end{tabu}
\end{table}

\subsubsection{Discussion}
The results support H4.1, that embodiment is negatively affected by simulation delays, which is in line with previous findings \cite{waltemate_impact_2016}. The descriptive results reveal that a jitter (in the applied spectrum) can result in a decreased perception of \textit{ownership} and \textit{agency} compared to an average latency of approximately $207$ ms, which emphasizes the negative impact of latency jitter. However, the total delay of about $353$ ms performed worse than the jitter condition, which quantifies the impact of jitter to some extent.
The correlations confirmed that the VEQ picks up modifications in movement synchronicity. 
Although we found expected correlations between the \textit{agency} factor and the additionally assessed \textit{agency} (KE) and \textit{ownership} (KE) measures, variance analysis revealed that the VEQ showed \textit{greater sensitivity} in the present scenario. This may result from the fact that the (KE) questions were developed for the RHI (i.e., physical world body part) scenarios. Although the results confirmed the expected correlations between \textit{ownership} and \textit{agency}, we could show that the \textit{change} factor was not affected by modifications in the motion dimension, and thus, the VEQ validly discriminates the factors to this regard. One limitation is that the agency control measure adapted from the (KE) questionnaire had overall low reliabilities, which is, in turn, a further argument for the proposed VEQ.
In conclusion, we found significant impacts of latency and jitter on ownership and agency, and could further quantify the impact of jitter. 

\section{General Discussion and Conclusion}
Overarching the individual findings of the studies, our high-level goal was the construction and validation of the Virtual Embodiment Questionnaire (VEQ). The proposed scale is not merely theoretical, but instead, its factors are statistically confirmed through rigorous scale development. The CFA confirmed the previously explored factor of the change in the perceived body schema \cite{roth_alpha_2017}, which could especially be relevant for VR applications in the context of therapy and disorder treatment \cite{piryankova_owning_2014}, as we assume it is a potential predecessor of the Proteus effect \cite{yee_proteus_2007}. Future research may investigate potential moderations, e.g., whether higher ownership fosters a change perception. Regarding the performance of the VEQ, acceptable to good \textbf{reliabilities} were confirmed with analyses of Cronbach's $\alpha$ throughout all studies. In contrast to previous component identifications \cite{longo_what_2008}, the VEQ focuses on virtual experiences. Further, the VEQ showed higher sensitivity to manipulations compared to a previous scale constructions \cite{kalckert_moving_2012} that aimed at assessing physical world RHI-like experiments. Compared to related measures for physical world experiments (KE) \cite{kalckert_moving_2012}, the VEQ reacted more \textbf{sensitive} to performed latency manipulations. 
The VEQ also reacted \textbf{validly and sensitively} in the regard that we could confirm and further specify the impact of immersion: immersion specifically addresses the perception of a change in the perceived body schema.

We further showed that the proposed measure to part correlates with related constructs of presence or self-presence but discriminates those from virtual embodiment components. By investigating internal correlations of the VEQ, we could also confirm a correlation between agency and ownership as mentioned in previous works (e.g., \cite{tsakiris_having_2006,braun_senses_2018}). The \textit{change} component assessing the perceived change of one's own body schema seems to be separable, in most regards. We interpret the confirmation of this component to be specifically relevant for VR. We confirmed that immersion (first-person perspective and HMD display) is a driver of this component, and thus desktop games or simulations or third-person VR simulations may not evoke this perception. 

Some limitations arise. In contrast to the suggestions by Gonzalez-Franco and Peck, the VEQ does not include ``external appearance,'' or ``response to external stimuli,'' which may be the subject of future work. However, as also noted by Gonzalez-Franco and Peck \cite{gonzalez-franco_avatar_2018} and reviewed in Section 1, the previous literature topics body ownership, agency, and self-location in the context of embodiment (e.g. \cite{kilteni_sense_2012,kalckert_moving_2012,piryankova_owning_2014,maselli_building_2013,blanke_multisensory_2012,blanke_full-body_2009}). The VEQ lacks of the assessment of self-location. The initial and updated questionnaire did not strictly concentrate on this factor. We argue that disturbed self-location (self-localization) is a factor typically present with disorders (paroxysmal illusions) evoking disrupted body perception, such as heautoscopy, out-of-body experiences, or autoscopic hallucinations (see \cite{braun_senses_2018,lopez_body_2008}). Although self-location discrepancies can be assessed in RHI experiments \cite{botvinick_rubber_1998} or out-of-body experiences \cite{aymerich-franch_embodiment_2015}, it is typically not the goal of VR applications to evoke such effects but rather to accurately reassemble the user's location, first-person perspective (see \cite{kilteni_sense_2012} for a discussion), and behavior. Therefore, self-location may be subject to future extensions or alternatively assessed with implicit measures, such as displacement measures \cite{botvinick_rubber_1998}. 
The present findings are also limited by the applied scenarios and the collected sample form and size. Future work should consequently assess the reliability and consistency of the VEQ, along with its application to different (non-mirror) scenarios, such as user-embodying games, and more abstract avatar types with varying form and appearances. In this regard, we can only assume that the measure is \textbf{objective} (i.e., shielded from third-party bias) on the basis of our results, as shown in a multi-study setting. The questionnaire is available at [url] in [languages]. The simulation codes are available upon request.

In conclusion, we presented the creation of a validated virtual embodiment questionnaire (VEQ) that can be applied to various VR experiments to assess latent factors of virtual embodiment.



\bibliographystyle{abbrv-doi}

\bibliography{daniel}
\end{document}